# Disentangling Proxies of Demographic Adjustments in Clinical Equations

Aashna P. Shah[1], James A. Diao[1,2], Emma Pierson[3], Chirag J. Patel[1&], Arjun K. Manrai[1&]

[1] Department of Biomedical Informatics, Harvard Medical School, Boston, MA
[2] Department of Medicine, Brigham and Women's Hospital, Boston, MA
[3] Department of Electrical Engineering and Computer Science, University of California, Berkeley, Berkeley, CA, USA

[&] Equal contribution
*Correspondence: Arjun_Manrai@hms.harvard.edu or 10 Shattuck St., Boston, MA, 02115

# Abstract

The use of coarse demographic adjustments in clinical equations has been increasingly scrutinized. In particular, adjustments for race have sparked significant debate with several medical professional societies recommending race-neutral equations in recent years. However, current approaches to remove race from clinical equations, including averaging race-specific equations or refitting without race, do not address the underlying causes of observed differences. Here, we present ARC (**A**pproach for identifying p**R**oxies of demographic **C**orrection), a framework to identify explanatory factors of group-level differences, which may inform the development of more accurate and precise clinical equations. We apply ARC to spirometry tests, ubiquitous physiological measures of pulmonary function that have traditionally been race-stratified, across two observational cohorts comprising 159,893 participants. Cross-sectional sociodemographic or exposure measures did not explain differences in reference lung function across race groups beyond those already explained by age, sex, and height. By contrast, sitting height accounted for up to 26% of the remaining population-level differences in lung volumes between healthy Black and White adults. We then demonstrate how pulmonary function test (PFT) reference equations can incorporate these individual-level factors in a new set of equations called $ARC_{PFT}$ that includes sitting height and waist circumference and, in both NHANES and UK Biobank, surpassed the predictive performance of recently introduced race-neutral GLI-Global equation recommended by major pulmonary societies. When compared to the GLI-Global method, inclusion of sitting height and waist circumference in $ARC_{PFT}$ decreased the mean absolute error by 13% among Black participants in the UK Biobank and by 24% in the National Health and Nutrition Examination Survey (NHANES). Furthermore, $ARC_{PFT}$ demonstrated reduced vulnerability to domain shift compared to race-based methods, with mean absolute error 19.3% and 35.6% lower than race-stratified models in out-of-sample Asian and Hispanic populations, respectively. This approach provides a promising path for understanding the proxies of imprecise demographic adjustments and developing personalized equations across clinical contexts.

# Introduction

In recent years, concern has grown over the use of demographic adjustments, in particular those based on race, in clinical algorithms including equations used to estimate kidney function, lung function, and the risk of other clinical outcomes[1–5]. Although there is broad consensus on the need to reconsider imprecise, group-level variables, considerable debate and uncertainty remain about how best to revise equations across different clinical contexts[6–11]. For example, several approaches to replace race in equations involve refitting the models without the race variable[7,9,12–15]. These methods, however, generally struggle to improve overall accuracy, overlook individual-level factors proxied by demographic variables, and may involve trade-offs in both performance and clinical implications[4,12].

To address these challenges, here, we present ARC (**A**pproach for identifying p**R**oxies of demographic **C**orrection), a systematic framework to identify the individual-level factors currently obscured by demographic adjustments in common clinical equations. We apply ARC to race adjustments in pulmonary function tests (PFTs), a domain complicated by the complex interplay of genetic, exposure, and social factors influencing lung function (Figure 1)[16–22]. In a typical spirometry evaluation, the patient will forcibly exhale into a device which measures total volume exhaled over time. Measured lung volumes, including the volume exhaled in one second (forced expiratory volume in 1 second—FEV1) or in one full breath (forced vital capacity—FVC), are then compared to a reference "normal" distribution calculated based on the patient's demographic features, which may include age, sex, height, and race[10,23–25]. Many clinical decisions, such as the diagnosis of chronic obstructive pulmonary disease (COPD) and restrictive diseases, are therefore based not on the directly measured lung volume, but on the percentage of predicted normal (percent-predicted) or by comparison to a lower limit of normal (LLN)[4].

Based on population-level data, Black participants were observed to have 10–15% smaller average lung capacity compared to White participants, and Asian participants were observed to have 4–6% smaller average lung capacity compared to White participants[26–28]. These differences were assumed to reflect biological variation between race groups; however, socioeconomic, environmental, and clinical factors that also influence physiology remain underexplored. Leveraging data from 159,893 individuals in two observational cohorts spanning from the US and UK, we we systematically examine the association of anthropometric, sociodemographic, and exposure variables to lung function across populations. Using these findings, we develop new equations, $ARC_{PFT}$, and evaluate their performance against both race-neutral and race-based approaches for predicting reference values and clinical outcomes.

## Results

**Overview of an Approach for Identifying Proxies of Race Correction (ARC) in Clinical Equations**

ARC (**A**pproach for identifying p**R**oxies of demographic **C**orrection) is a framework for identifying individual-level factors obscured by coarse demographic variables in clinical equations. ARC involves five key steps (Figure 1): (1) assemble reference cohorts, (2) select outcome-relevant variables, (3) identify variables explaining population-level differences (e.g., race, age, geographic location), (4) develop models excluding each demographic variable systematically, and (5) evaluate these models against demographically-adjusted models across in- and out-of-distribution datasets. While we apply ARC to race correction in lung function equations as a case example, the framework is broadly applicable to other demographic adjustments commonly used in clinical algorithms.

**Study population and race-adjusted differences in pulmonary function tests**

We used two cohorts, UK Biobank and NHANES, with geographic and temporal differences to assess the generalizability of reference equations and the consistency of associations across cohorts. The reference cohort consisted of adults without recent respiratory symptoms or history of smoking or lung disease between ages of 40 and 80 years old, including 156,526 participants from UK Biobank and 3,367 participants from NHANES III-IV. The UK Biobank cohort consisted of 94.9% White, 1.2% Black, 2.2% Asian, and 1.5% individuals of other or unknown ethnicities (Table 1). The NHANES cohort consisted of 39% White, 22% Black, 3% Asian, and 36% from other ethnic backgrounds.

Differences across race groups in FVC and FEV1 were observed in both cohorts after adjusting for sex and age (Figure 2, Figure S1, Table S2-S5). Specifically, compared to White individuals, Black individuals exhibited an FVC difference of -871 ± 14 mL in UK Biobank and -687 ± 27 mL in NHANES. Asian individuals showed differences of -1,014 ± 11 mL and -867 ± 65 mL, respectively[18]. Hispanic individuals in NHANES had an FVC difference of -417 ± 25 mL. Participants of other or unknown race showed differences of -445 ± 13 mL and -695 ± 51 mL in UK Biobank and NHANES, respectively.

**Domain-wide association between FVC and anthropometric, sociodemographic, and exposure variables**

We assessed whether the observed population differences in lung function could be explained by anthropometric, sociodemographic, and exposure variables. Specifically, we collected anthropometric (e.g., sitting height, standing height, weight), social (e.g., income, education, immigration status), and exposure (smoke exposure via dust or fumes, particulate matter, etc.) variables with prior evidence of mechanistic influences on health outcomes[16–19], confirmed associations across cohorts, and quantified the ability of these variables to explain population differences in lung function (Figures 2-3, Figures S1-S3).

In the UK Biobank, all 15 variables tested were significantly correlated with lung function after controlling for age, sex, and age-sex interaction. In the NHANES dataset, 21 of 23 variables tested showed significant correlations with lung function. Anthropometric variables such as standing height, sitting height, and weight demonstrated strong positive correlations in both cohorts, while BMI and waist circumference showed negative associations. Education, income level, and immigration status exhibited positive correlations with FVC in both UK Biobank and NHANES. Exposure to household smoke and fine particulate matter (PM2.5) were negatively associated with FVC in both the UK Biobank and NHANES cohorts, with consistent direction and magnitude of associations across cohorts.

**Anthropometrics explain a substantial proportion of observed population differences in pulmonary function**

To systematically quantify the association of individual-level factors with population differences in lung function across race groups, a series of regression analyses were conducted. The initial model included race, age, sex, and the age-sex interaction (FVC ~ race + age + sex + age·sex). Then, variables that were significant from the domain-wide association study were incrementally added to the model. We then measured the change in the race coefficient, defined with reference to White individuals, upon inclusion of each variable.

In UK Biobank, height explained 39% and 43% of the FVC difference for Asian and Other groups, respectively, compared to White individuals, but only 17% in the Black subgroup (Figure 3a). Similarly, in NHANES, height accounted for 60% (Asian), 61% (Other), but just 5% (Black) of the FVC differences (Figure 3b). Notably, adjusting for height reduced the difference between Hispanic and White groups by 126%, reversing the direction of the association from negative to positive.

Even after including height, other anthropometric variables accounted for significant differences across race groups in FVC and FEV1 (Figure 3, Figures S3-S5, Tables S2-S5). In UK Biobank, sitting height explained an additional 8% of the FVC difference in Black individuals, while it explained approximately 3-4% in Asian and Other populations (Figure 3a). Similarly, in NHANES, sitting height contributed to 26% of the FVC differences for the Black population, and 9%, 12%, and 5% in Asian, Hispanic, and Other populations, respectively (Figure 3b). Comparable results were observed for $FEV_{1.}$.

After adjusting for sex, age, height, and sitting height, other sociodemographic variables and exposures were found to have minimal association with race-related differences in FVC and FEV1 (Figure 3, Figures S3-S5). Even after excluding anthropometric variables, exposures failed to explain race-related differences, despite showing significant correlations with lung function (Tables S10-S13). By contrast, sociodemographic factors explained 6-8% of the race-related difference in lung function in the UK Biobank cohort and 10-38% of the difference in the NHANES cohort (Table S6-S9).

**Anthropometrics, not sociodemographics, improve FVC prediction across cohorts**

We refit the GLI-Global 2022 equation using UK Biobank data and developed a comparable baseline model, $ARC_{PFT}$, excluding group-level weighting to assess the impact of inverse probability weighting (Methods). Both models were evaluated on an internal test set of 31,306 individuals from the UK Biobank. $ARC_{PFT}$ demonstrated comparable performance to the refit GLI-Global 2022, with the macro-mean absolute error (MAE), the average MAE across each race, showing no significant differences. In the UK Biobank test set, the macro-MAE was 0.60 ± 0.17 L for the GLI-Global 2022 model, while it was 0.59 ± 0.16 L for $ARC_{PFT}$ (Figure 4a, Table S14). External validation in the NHANES cohort (n = 3,367; 39% White, 22% Black, 3% Asian, 36% Other) exhibited comparable performance, with a macro-MAE of 0.47 ± 0.09 L for both the refit GLI-Global 2022 and $ARC_{PFT}$ models (Figure 4b, Table S16).

We tested whether $ARC_{PFT}$ could be improved with additional anthropometric (e.g., sitting height, waist circumference) and sociodemographic variables (e.g., immigration status), finding that both substantially improved predictive performance for Black and Asian individuals, while having minimal effect for White and Other populations. For Black and Asian individuals, the baseline MAE (Black: 0.75 L; Asian: 0.70 L) decreased progressively with the inclusion of sitting height (Black: 0.68 L; Asian: 0.65 L), waist circumference (Black: 0.65 L; Asian: 0.64 L), and immigration status (Black: 0.51 L; Asian: 0.46 L). In contrast, the addition of smoke exposure led to negligible performance gains across all groups (Figure 4a, Table S14).

Consistent with the UK Biobank findings, adding anthropometric and sociodemographic variables did not significantly improve performance for White (MAE: 0.41 L, 95% CI: 0.39–0.42) or Other (MAE: 0.49 L, 95% CI: 0.43–0.54) individuals in the NHANES cohort. However, for Black and Asian individuals, the MAE for $ARC_{PFT}$ (Black: 0.62 L; Asian: 0.47 L) progressively decreased with the inclusion of sitting height (Black: 0.55 L; Asian: 0.44 L) and waist circumference (Black: 0.47 L; Asian: 0.40 L). For Hispanic individuals, the MAE for $ARC_{PFT}$ (0.38 L) did not significantly change after the addition of further anthropometric variables.

In contrast, in NHANES, incorporating sociodemographic variables often had negligible or even negative effects on performance. For example, adding immigration status had little impact on MAE when anthropometric variables were already included (Black: 0.48 L; Asian: 0.40 L). Notably, for Hispanic individuals, the MAE increased from 0.38 L to 0.55 L with the inclusion of immigration status, indicating possible sensitivity to cohort differences or domain shift (Figure 4b, Table S16).

**Race adjustment and performance in Black individuals**

For each race-neutral $ARC_{PFT}$ model, we compared its performance to a race-adjusted counterpart by incorporating race as a feature. While the impact of incorporating race on White individuals was minimal, with mean absolute error (MAE) reductions ranging from 0.1% to 3.0%, it significantly increased performance for Black individuals. In the UK Biobank, including race reduced the MAE for Black individuals by 47.4% when height was included, 42.6% with the addition of sitting height, 40.3% with waist circumference, and 24.8% with immigration status (Figure 4a, Table S15). Notably, this impact of race adjustment diminished as more demographic and anthropometric variables were added, suggesting that the models became less reliant on race as covariates were included.

Similarly, in the NHANES cohort, incorporating race reduced the MAE for Black individuals by 29.5% when height was included, 20.5% with sitting height, and 0.9% with waist circumference. The MAE decreased by 10.7% with the incorporation of immigration status (Figure 4b, Table S16). Comparable improvements were observed when smoke exposure was included, and similar trends were noted for FEV1 predictions (Figure S3, Table S18, S20).

**$ARC_{PFT}$ outperforms race-adjusted models in generalizing to Asian and Hispanic populations**

For Asian populations, race adjustment decreased the MAE in the UK Biobank but often increased it in the out-of-distribution NHANES test set. In the UK Biobank, incorporating race reduced the MAE for Asian individuals by 48.3% when height was included, 44.5% with the addition of sitting height, 45.6% with waist circumference, and 23.3% with immigration status (Figure 4a, Table S15). In contrast, in NHANES, incorporating race increased the MAE by 1.8% with height, 9.3% with sitting height, 6.7% with waist circumference, and 19.3% with immigration status (Figure 4, Figure S3, Table S16, Table S17).

Similarly, race-neutral equations outperformed race-adjusted models in the NHANES Hispanic population (Figure 4b, Figure S3b, Table S16, S17). Hispanic ethnicity was not explicitly defined in the UK Biobank, so NHANES was used to assess the robustness of race-adjusted equations on out-of-distribution groups. Since the race-adjusted models lacked a specific category for Hispanic individuals, individuals were classified as Other. Compared to the race-neutral models, incorporating race increased the MAE by 22.1% with Height, 23.8% with sitting height, 35.6% with waist circumference, and 4.0% with immigration status.

## Discussion

Here, we introduce the ARC framework to uncover individual-level drivers of population-level differences that may be proxied by coarse demographic adjustments in clinical equations. In particular, we apply ARC to disentangle the proxies of race adjustments in common estimates of lung function. Race has historically been included as a predictor of lung function, and ARC enabled us to identify key factors that explain a substantial portion of population differences in lung capacity[23,24]. By pinpointing these factors, we highlight the limitations of existing models and provide guidance for developing more precise alternatives.

Current race-neutral pulmonary function models rely on height, sex, and age as covariates, but our findings show that height alone is insufficient to capture race-related differences in lung capacity. While height explained nearly all differences in FVC among Hispanic individuals, it explained only a small portion of the variation in FVC among Black individuals. Sitting height explained additional differences in lung function not explained by standing height, consistent with prior studies[14]. However, observed differences require further investigation of social and environmental contributors. While earlier studies suggested only modest gains from including sitting height, our analysis, using larger datasets and more flexible modeling, found greater improvements, particularly among Black individuals[14]. Furthermore, waist circumference also improved predictive accuracy among Black individuals, but its non-linear relationship with lung function and time-varying nature may limit its suitability for predicting baseline lung function[29,30].

Sociodemographic factors, including immigration status, income, and education level also improved predictive accuracy for Black and Asian populations within the UK Biobank. However, evaluation on the out-of-sample NHANES cohort revealed that incorporating these cross-sectional sociodemographic variables did not improve performance for most groups and, in fact, significantly reduced accuracy for Hispanic populations, likely reflecting a differences between the two cohorts.

Our results also showed that, within the reference populations, readily measurable exposures did not account for observed differences in lung function across race groups. This may reflect that the known effects of environmental and socioeconomic exposures on lung health are not well captured by an individual's current status alone but rather by cumulative exposure over time, and especially during the childhood and adolescent phases of lung development[31–33]. Moreover, some exposure-related effects may already be partially reflected in anthropometric measurements, which integrate aspects of long-term growth and development[34]. Future work should aim to characterize and quantify these longitudinal and interrelated factors. However, our findings suggest that current efforts to incorporate readily measurable exposures and sociodemographic variables may not capture the true effects of lifelong exposure on health[35].

We used two large, population-based cohorts, UK Biobank and NHANES, which differ in geographic context, time period, population demographics, and data collection methods. This enabled us to evaluate the robustness of demographic correction approaches and underscores the importance of assessing clinical equations across heterogeneous datasets. Our findings highlight that the effects of race adjustment are highly sensitive to inter-cohort differences, particularly those stemming from sociodemographic variation[36–38] and inconsistencies in race and ethnic categorization[39–42]. These results underscore the need for caution when using coarse group-level variables like race in predictive models, as such categories evolve over time and vary across contexts[42,43]. The performance of race-neutral models in external validation demonstrates their potential for generalizability and for better capturing the complexity of pulmonary function across diverse populations, compared to race-adjusted or inverse-weighting approaches. When variation associated with group-level variables can be explained by more specific and stable individual-level factors, those factors should be considered to improve both the precision and robustness of clinical algorithms.

Our study has several limitations. Evaluation metrics comparing predicted to observed values in reference cohorts can be misleading if reference cohorts inadvertently include individuals with subclinical or underreported disease, leading to overfitting and the normalization of poorer health outcomes. Further studies should investigate associations with concurrent or incident outcomes of direct relevance to patients, including symptoms, hospitalizations, new-onset respiratory disease, and mortality, as performed by prior studies comparing race-neutral and race-adjusted GLI equations[4]. Furthermore, while incorporating race in $ARC_{PFT}$ improved performance for Black individuals, other factors not interrogated in this study–including genetic predispositions, early-life and adolescent exposures, and additional environmental factors–likely contribute to lung function disparities and may help explain the remaining differences[16,31–33,44–47]. Finally, our analyses used broad race categories, which may mask important within-group heterogeneity. More granular classifications could uncover additional performance differences and further inform model development.

While demographic correction such as race has been used in clinical equations, its use remains controversial and context-dependent[1,11,48,49]. Such adjustments can enhance predictive accuracy in some settings, but they may also mask underlying biological, environmental, or structural factors driving group differences. By systematically uncovering the specific factors that traditional models capture through demographic proxies, ARC and related methods offer a path toward more precise, individualized, and robust clinical algorithms.

## Methods

### Study Population

This study utilized spirometry data collected from adult participants in the UK Biobank (2006-2010) and the NHANES (1988-1994; 2007-2008) cohorts. The breath spirometry was collected using a Vitalograph Pneumotrac 680 and dry-rolling seal volume spirometers for the UK Biobank and NHANES cohorts, respectively. Participants who did not meet the acceptability and reproducibility criteria set by the ATS were excluded from the study. We identified spirometry measurements as those passing the 2005 ATS technical standard for interpreting spirometry with the maximum $FEV_1$ and FVC values from at least three acceptable curves[50].

A reference cohort was created by applying Hankinson and colleagues' criteria, which focused on asymptomatic, lifelong nonsmoking adults ranging in age from 40 to 80 years [23]. Individuals with lung disease (asthma, COPD, pulmonary fibrosis, bronchitis, asbestosis, emphysema, tuberculosis, cancer), lung symptoms (coughing, wheezing, or difficulty breathing), or an $FEV_1/FVC < 0.7$ during previous or future visits were excluded. Furthermore, participants included complete anthropometric, sociodemographic data (age, sex, self-identified race, income, and education), and exposure (smoke) information. Missing self-reported race data was categorized as Other, while missing clinical and exposure data were assumed to indicate absence of disease, symptoms, or exposures (i.e., coded as 'False').

### Imputation of Sitting Height

To address missing sitting height in NHANES IV (2007-2008), we fit an eXtreme Gradient Boosting (XGBoost) model to predict sitting height using sex, age, height, weight, body mass index, waist circumference, and race[51]. The training set was made through an 80:20 split using 494,980 participants from UK Biobank (395,984 for the development set and 98,996 for the test set). Hyperparameters (learning rate, max depth, number of estimators) were selected using three-fold cross-validation on an 80:20 split of the development set (316,787 for each training set and 79,197 for validation set). The model was validated using mean absolute error on the 98,996 participant test set in the UK Biobank and 16,787 participants from NHANES III cohort, and was employed on NHANES IV (Table S1).

**Domain-Wide Association Study**

For each cohort, we conducted a domain-wide association study to assess the relationship between sociodemographic (n=15), anthropometric (n=9), and exposure (n=3) variables with $FEV_1$ and FVC. Numerical covariates were summarized using standardized mean differences. White individuals were the reference population when assessing the association between race and spirometry measures. The association for each covariate was calculated, adjusting for sex, age, and the age-sex interaction. To account for multiple testing (UK Biobank: n = 15, NHANES: n = 23), two-sided p-values were corrected using the Bonferroni method[52].

**Quantifying Associations with Population Differences**

To evaluate the association of anthropometric, socioeconomic, and exposure variables on race-related differences in lung function, we performed a series of regression analyses in each cohort, adjusting for age, sex, and their interaction. Covariates were added iteratively using a forward selection approach, with inclusion determined by the greatest reduction in mean squared error (MSE). Continuous variables were modeled using cubic splines with four degrees of freedom.

**Developing PFT Equations**

The GLI-Global 2022 equation, the current race-neutral standard, eliminates the use of race from the previous GLI-2012 model by applying inverse probability weights for ethnicity and sex, while retaining height, age, and sex as predictors of reference $FEV_1$ and FVC[14]. We replicated this training procedure using data from 125,220 UK Biobank participants (5% White, 1% Black, 2% Asian, 2% Other) to produce a GLI-Global 2022–style model for comparison.

To isolate the effect of inverse probability weighting, we developed a baseline model, termed $ARC_{PFT}$, using the same predictors as GLI-Global 2022 but without group-level weights. $ARC_{PFT}$ was trained using XGBoost, allowing for the flexible inclusion of additional covariates such as waist circumference, sitting height, education, immigration status, and smoke exposure[51]. Covariates were added in a stepwise manner and hyperparameters (learning rate, max depth, number of estimators) were selected using three-fold cross-validation. The models were validated using the UK Biobank test set and NHANES dataset. For each race-neutral model, a race-adjusted model was created by including race as a covariate. The performance of the models was evaluated using mean absolute error (MAE).

**Data Availability**

This research was conducted using data from the UK Biobank (Application Number 22881). UK Biobank data are available by application via https://www.ukbiobank.ac.uk/. The National Health and Nutrition Examination Survey (NHANES) data used in this study are publicly available and can be accessed from the Centers for Disease Control and Prevention (CDC) website: https://www.cdc.gov/nchs/nhanes/.

**Code Availability**

The code supporting the findings of this study is available at https://github.com/aashnapshah/arc_pft.

# Figures

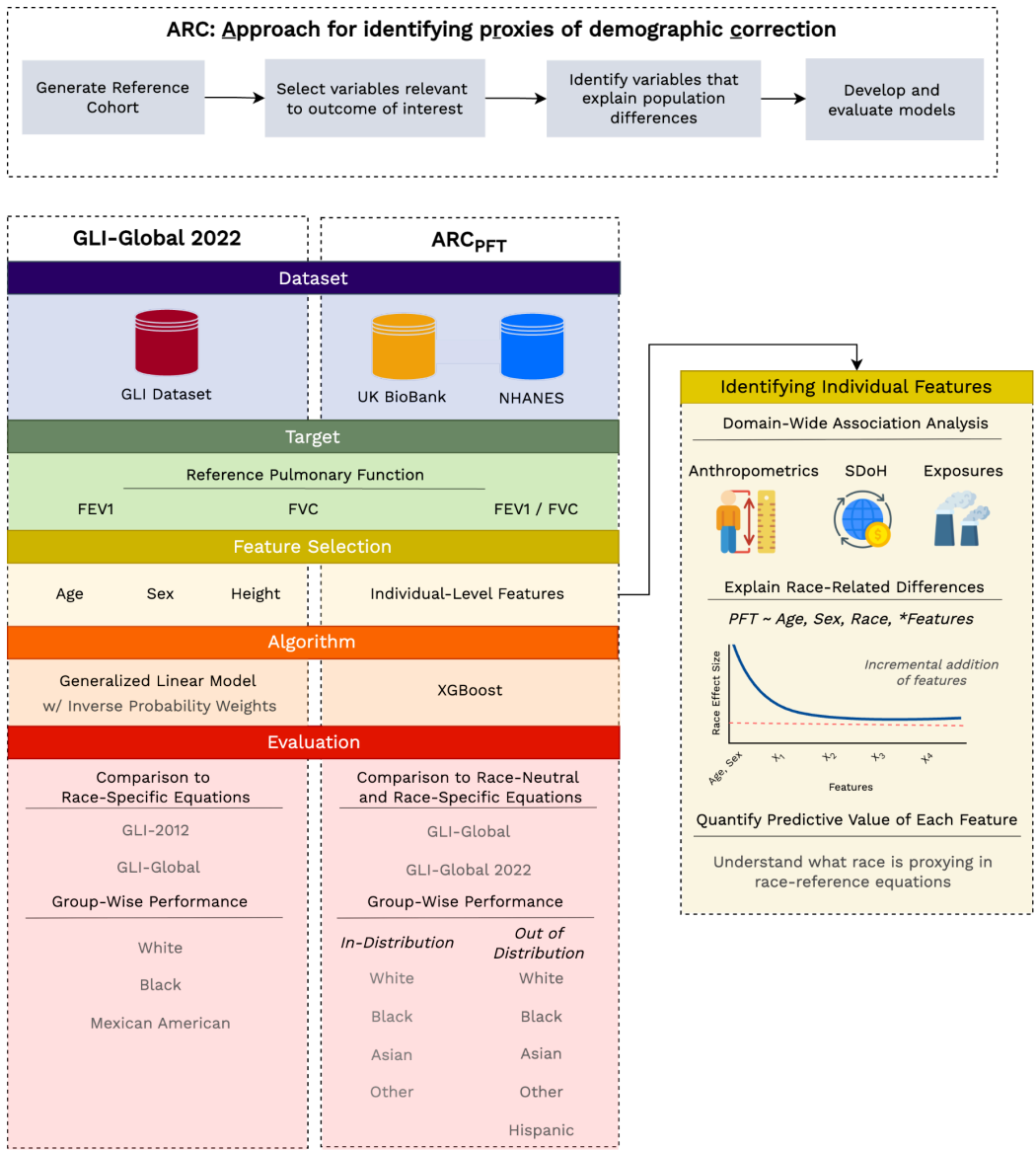

**Figure 1. Overview of the GLI-Global and ARC$_{PFT}$ frameworks for disentangling the use of race in reference pulmonary function test equations.** The GLI-Global equation leveraged the GLI dataset to model reference pulmonary function (FEV1, FVC) using generalized linear models with age, sex, and height as predictors. This equation was evaluated against race-adjusted equations (GLI-2012) and assessed for group-wise performance across White, Black, and Mexican American populations. The ARC$_{PFT}$ equation utilized UK Biobank and NHANES datasets, employing XGBoost algorithms with individual-level features which explain race-related differences. The framework compared race-neutral (GLI-Global) and race-specific equations (GLI-2012) and evaluated groupwise performance both in-distribution and out-of-distribution for diverse race groups.

**Table 1. Study Population Demographic and Anthropometric Measurements of UK Biobank and NHANES Cohorts.**
Race and ethnicity were self-reported in both cohorts. UK Biobank does not include individuals identifying as Hispanic, and the Other category includes multiracial or other unspecified ethnicities. In NHANES, White, Black, and Asian categories represent non-Hispanic individuals, with Hispanic individuals categorized separately. Abbreviations: FEV1 = forced expiratory volume in 1 second (L), FVC = forced vital capacity (L), FEV1/FVC = ratio of FEV1 to FVC, SD = standard deviation, Waist Circ. = waist circumference, cm = centimeters. Education refers to the percentage of participants who completed high school.

|  | UK BioBank | | | | NHANES | | | | |
| --- | --- | --- | --- | --- | --- | --- | --- | --- | --- |
|  | White | Black | Asian | Other | White | Black | Hispanic | Asian | Other |
| **Sample Size** | 148,541 | 2,029 | 3,504 | 2,452 | 1337 | 739 | 1,053 | 89 | 149 |
| **Female (%)** | 59% | 61% | 55% | 59% | 63% | 65% | 64% | 63% | 64% |
| **Age (SD) - years [1]** | 57.2 (7.9) | 52.3 (7.9) | 53.8 (8.2) | 53.7 (8.1) | 57.9 (12.0) | 54.3 (11.0) | 54.0 (10.5) | 53.7 (9.9) | 52.9 (10.1) |
| **FEV1 (SD) - L[1]** | 2.9 (0.7) | 2.4 (0.6) | 2.3 (0.6) | 2.7 (0.7) | 2.9 (0.8) | 2.5 (0.7) | 2.8 (0.7) | 2.4 (0.7) | 2.6 (0.7) |
| **FVC (SD) - L[1]** | 3.7 (0.9) | 3.0 (0.8) | 2.9 (0.8) | 3.4 (0.9) | 3.8 (1.0) | 3.2 (0.9) | 3.5 (0.9) | 3.0 (0.9) | 3.2 (0.9) |
| **$FEV_1$/FVC (SD) [1]** | 0.8 (0.1) | 0.8 (0.1) | 0.8 (0.1) | 0.8 (0.1) | 0.8 (0.1) | 0.8 (0.1) | 0.8 (0.0) | 0.8 (0.1) | 0.8 (0.0) |
| **Height (SD) - cm** | 168.1 (9.0) | 166.7 (8.4) | 162.9 (8.8) | 165.8 (9.2) | 166.7 (9.8) | 167.1 (9.0) | 160.5 (9.1) | 160.0 (8.7) | 161.5 (8.8) |
| **Sitting Height (SD) - cm [1]** | 89.1 (4.7) | 85.8 (4.5) | 85.4 (4.8) | 87.5 (5.0) | 87.9 (4.9) | 85.6 (4.3) | 84.6 (4.6) | 83.9 (4.3) | 85.1 (4.8) |
| **Waist Circ. (SD) - cm[1]** | 88.7 (13.0) | 91.3 (11.9) | 88.3 (12.1) | 88.8 (12.9) | 97.3 (14.5) | 100.3 (14.7) | 97.7 (11.8) | 87.5 (9.7) | 90.9 (13.8) |
| **Immigrant (%)** | 4% | 70% | 89% | 50% | 6% | 15% | 62% | 97.8% | 85% |
| **Education (%)** | 56% | 63% | 64% | 66% | 65% | 53% | 35% | 76% | 57% |
| **Smoke Exposure (%)** | 11% | 3.4% | 3.5% | 7.0% | 7.9% | 14% | 8.3% | 2.2% | 9.4% |

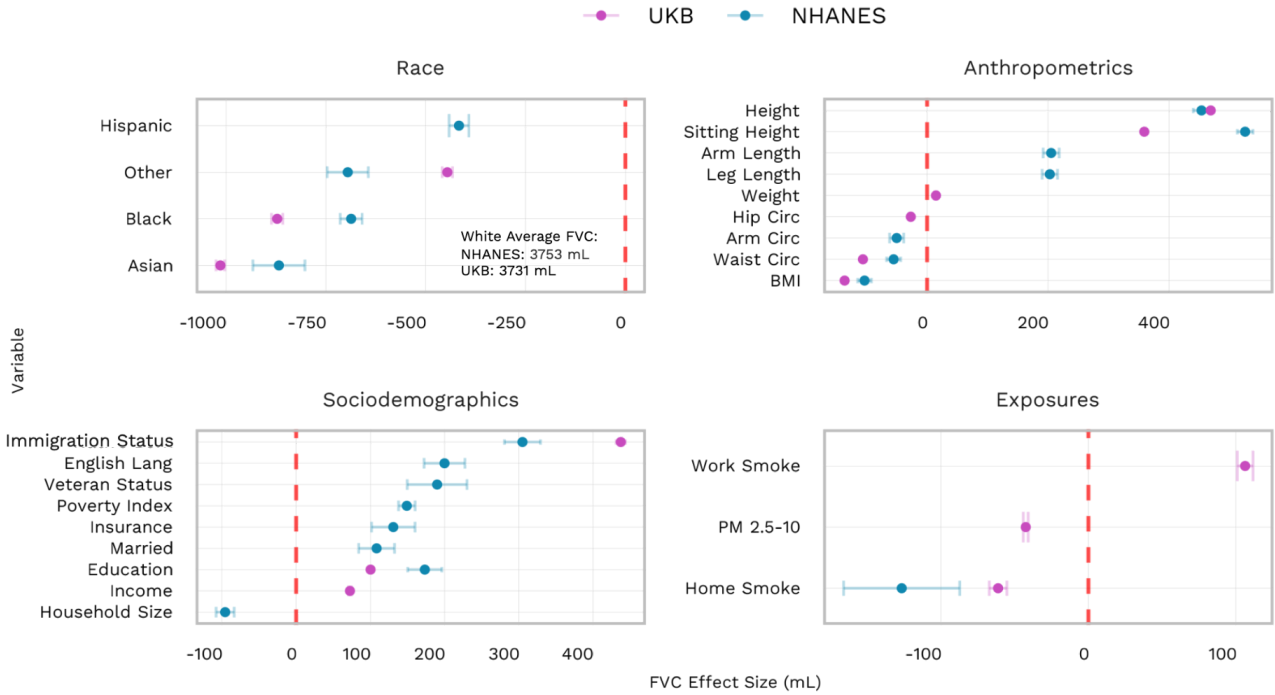

**Figure 2. Association of race, anthropometrics, sociodemographics, and exposures on FVC (mL)** For each cohort, the association of significantly associated variables, including race, anthropometrics, sociodemographics, and exposures on FVC, is displayed. Each model was adjusted for sex, age, and their interaction (sex*age). White individuals were treated as the reference group, and associations for sociodemographics were evaluated against the following reference categories: completed high school for education, lowest income level (0) for income level, and English speakers for language. Continuous variables were standardized using their mean and standard deviation (SD), and associations are expressed in units of SD. Error bars represent the standard deviation of the association estimate. Abbreviations used in the figure include circumference (circ.), body mass index (BMI), language (lang.), education refers to finishing high school, income level is categorized as 0 (lowest), 1, 2, and 3 (highest), and particulate matter (PM).

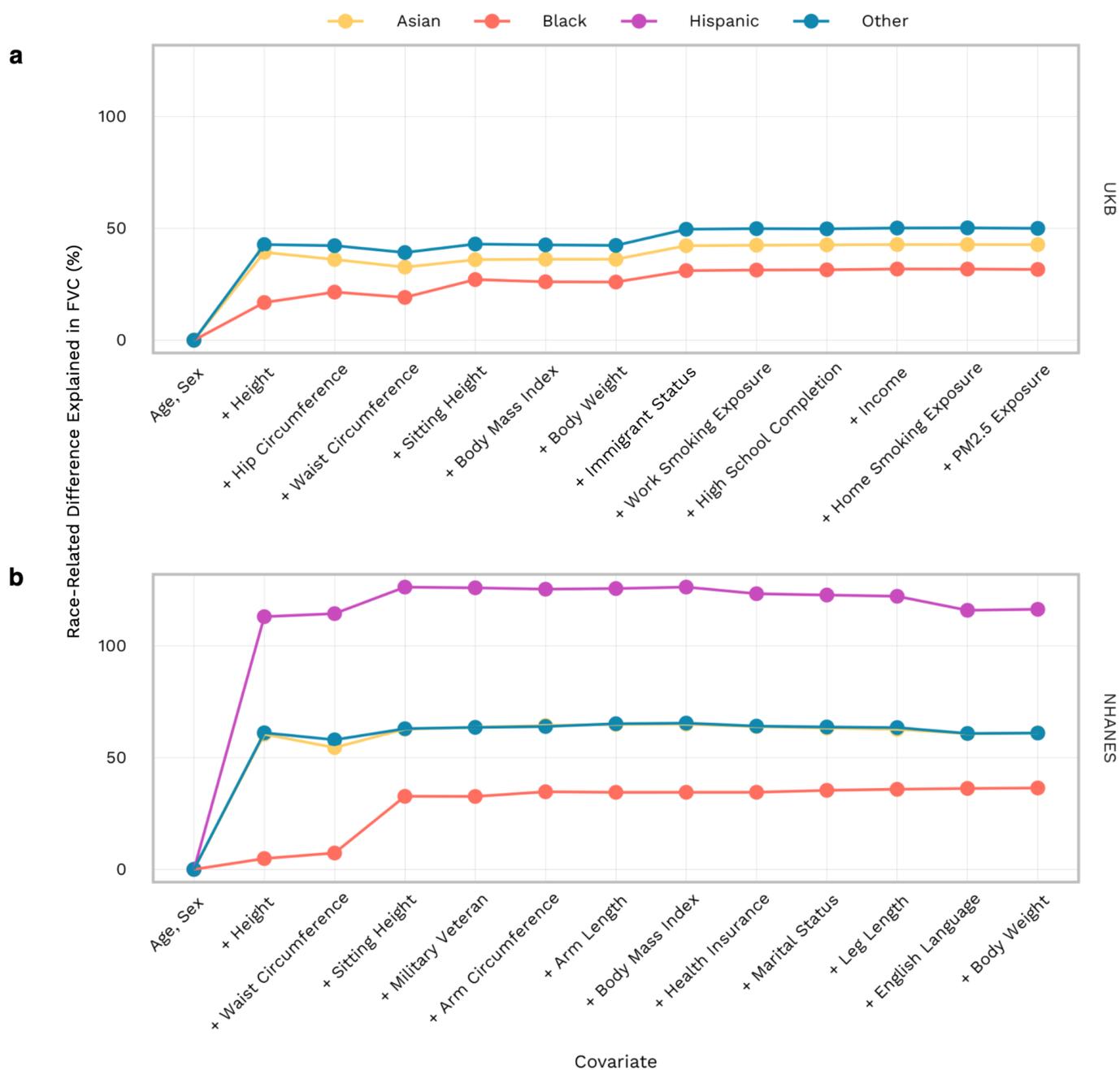

**Figure 3. Race-related differences explained by anthropometrics, sociodemographic, exposure variables on the FVC in healthy, asymptomatic individuals.** For the UK Biobank and NHANES cohorts, the proportion of race-related differences in FVC (adjusted for age and sex) explained by stepwise inclusion of individual-level variables is shown. Cubic splines were applied to numerical values to account for potential non-linear relationships.

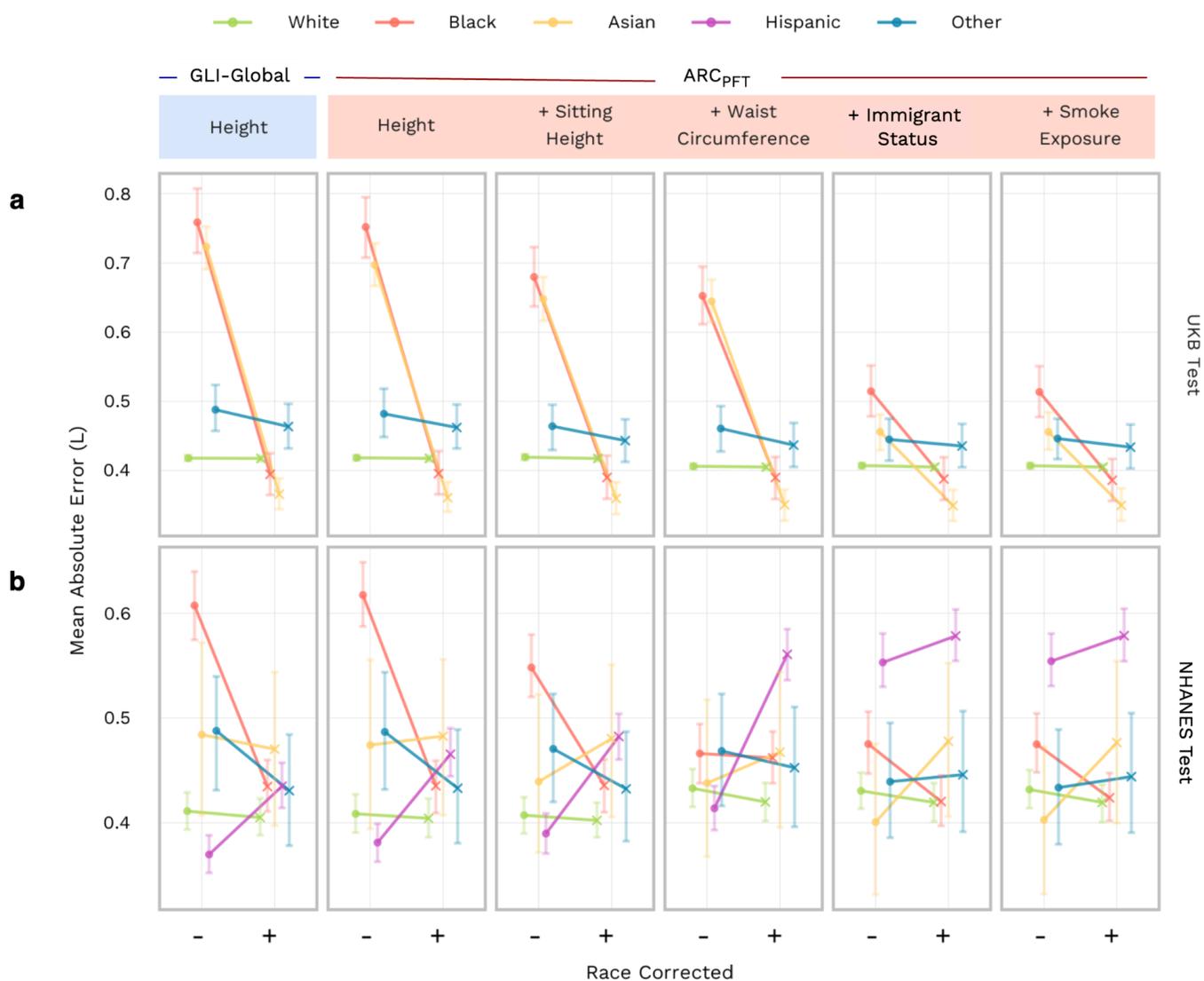

**Figure 4. Performance of GLI-Global and ARC$_{PFT}$ models on reference FVC prediction.** Each model was trained using a subset of the UK Biobank data, with age, sex, and height as predictors. Sitting height, waist circumference, immigration status, and smoke exposure were incrementally added to the models. The models were evaluated using mean absolute error (MAE) on: (a) the UK Biobank test set and (b) the NHANES dataset. In race-based models, Hispanic was encoded as Other due to the absence of labeled Hispanic individuals in the UK Biobank training set.